\documentclass[prl,reprint, twocolumn,showpacs,preprintnumbers,amsmath,amssymb,nofootinbib,floatfix]{revtex4-1} 
\usepackage{graphicx}
\usepackage{mathrsfs}
\usepackage{hyperref}
\usepackage{slashed}
\usepackage{color}
\usepackage{booktabs}
\usepackage{ulem}

\newcommand{\beq}{\begin{eqnarray}}
\newcommand{\eeq}{\end{eqnarray}}

\newcommand{\zsl}{ z \hspace{-2truemm}/ }

\begin{document}

\title{Tomography of high-twist proton structure through $ep$ elastic scattering}

\author{Ji-Xin Yu$^1$}
\author{Shan Cheng$^2$}\email{Corresponding author: scheng@hnu.edu.cn}
\author{Jia-Jie Han$^1$}\email{Corresponding author: hjj@lzu.edu.cn}
\author{Hsiang-nan Li$^3$}\email{Corresponding author: hnli@phys.sinica.edu.tw}
\author{Fu-Sheng Yu$^1$}\email{Corresponding author: yufsh@lzu.edu.cn}

\affiliation{$^1$ 
Frontiers Science Center for Rare Isotopes and School of Nuclear Science and Technology, Lanzhou University, Lanzhou 730000, China}
\affiliation{$^2$ 
School of Physics and Electronics, School for Theoretical Physics and Hunan Provincial Key Laboratory of High-Energy Scale Physics and Applications, Hunan University, 410082 Changsha, China}
\affiliation{$^3$ Institute of Physics, Academia Sinica, Taipei, Taiwan 115, Republic of China}

\date{\today}

\begin{abstract}

We present the first high-twist study of the proton form factors $F_{1,2}(Q^2)$ in $ep$ elastic scattering based on the perturbative QCD $k_T$ factorization, 
$Q^2$ being momentum transfer squared. It is motivated by unexpectedly large higher-power contributions from subleading-twist 
light-cone distribution amplitudes (LCDAs), which are attributed to the enhancement in endpoint regions of parton momentum fractions. 
We highlight that the endpoint enhancement, tamed by the $k_T$ resummation effect, 
is crucial for accommodating the approximate scaling behavior of the  $Q^4F_1(Q^2)$ data at intermediate $Q^2\sim {\cal O}(10)$ GeV$^2$.
The proton LCDAs up to twist 6 are then extracted, and verified by the charge-parity asymmetries observed in hadronic heavy baryon decays.
Our work provides new insights into the proton three-dimensional structure and manifests the precision requirement for reliable perturbative analyses of baryonic exclusive processes.
  
\end{abstract}


\maketitle

\textbf{\textit{Introduction}}--The proton structure bridges fundamental physics for quark-gluon interactions at a femtometer scale and  nucleosynthesis at a cosmological scale. Its emergent phenomena challenge naive quark-model expectations, most notably for the origin of the proton spin, to which quark spins contribute only $\sim 30\%$ \cite{EuropeanMuon:1987isl,Kuhn:2008sy,Ji:2020ena,Yang:2016plb}, and for the proton mass decomposition \cite{Ji:1994av,Lorce:2017xzd,Yang:2018nqn} that remains ambiguous. 
The nucleon tomography \cite{Pohl:2010zza,Antognini:2013txn,Gao:2021sml,Fischer:2004jt,Karshenboim:2005iy} is pioneered by measuring  electromagnetic form factors (FFs) through electron-nucleon elastic scattering \cite{Hofstadter:1953zjy,Hofstadter:1956qs}. 
Theoretical investigations of FFs transit from phenomenological descriptions to ab initio calculations based on microscopic quark-gluon pictures 
with the advent of the QCD factorization (QCDF) \cite{Lepage:1979za,Chernyak:1977fk}. 
A fast moving proton in a hard process is predominantly characterized by a state of three collimated valence quarks in the infinite-momentum frame, 
which naturally introduces the concept of light-cone distribution amplitudes (LCDAs). 
Nonperturbative QCD dynamics responsible for the proton structure is encoded into LCDAs. 
The hard dynamics is captured by the subprocess with a minimum of two virtual gluon exchanges among the three valence quarks. 
The QCDF, formulated as convolutions of hard kernels with LCDAs in parton momentum fractions \cite{Chernyak:1984bm,Ji:1986uh,Carlson:1987sw,Thomson:2006ny}, predicts the  asymptotic behavior $\alpha^2_s(Q^2)/Q^4$ of the proton FFs  \cite{Brodsky:1973kr,Mueller:1981sg,SL-pQCD}, $\alpha_s$ being the strong coupling and $Q^2$ being momentum transfer squared.  

Two independent groups have computed the next-to-leading-order (NLO) contributions to the proton Dirac FF $F_1(Q^2)$ at leading-twist level in the QCDF recently \cite{Huang:2024ugd,Chen:2024fhj}.
Their results, despite being compatible with the data for some model LCDAs within sizable theoretical uncertainties,
fail to accommodate the observed approximate scaling of $Q^4F_1(Q^2)$ at intermediate $Q^2\sim {\cal O}(10)$ GeV$^2$. The calculations also reveal substantial radiative corrections, particularly in the above range of $Q^2$. 
This feature is attributed to the logarithmic enhancement around the endpoints of parton momentum fractions, which becomes significant as $Q^2$ decreases. 
The findings in \cite{Huang:2024ugd,Chen:2024fhj} thus motivate a more sophisticated handling of 
the contributions from the endpoint region and from subleading twists of LCDAs.
The endpoint enhancement implies that parton transverse momenta $k_T$ are relevant degrees of freedom in addition to momentum fractions.
The perturbative QCD (pQCD) approach based on $k_T$ factorization resumes important $k_T$-dependent logarithms near the endpoints 
to all orders in $\alpha_s$ \cite{Li:1992ce}, establishing a rigorous formalism for exclusive processes at intermediate scales \cite{Chai:2024tss} 
accessible in current and forthcoming experiments. 

We present the first pQCD study of the proton FFs at subleading powers, including the LCDAs up to twist 6 systematically. 
It will be shown that the endpoint enhancement maximizes the contributions to the FFs at combined twist 8, 
and the power expansion starts to converge above combined twist 9. 
The match of the pQCD predictions to the precise FF data from SLAC \cite{Arnold:1986nq,Sill:1992qw,Rock:1991jy} and JLab \cite{Christy:2021snt} 
selects the preferred parameters in the proton LCDAs, which confront the outcomes from QCD sum rules (QCDSR) \cite{Chernyak:1987nt}, 
lattice QCD (LQCD) \cite{RQCD:2019hps} and data-driven light-cone sum rules \cite{Anikin:2013aka}. 
The determined proton LCDAs, which ought to be universal, are verified by the improved consistency between the predictions and the data for the charge-parity violation in the heavy baryon decay $\Lambda_b \to pK$ \cite{Han:2024kgz,Han:2025tvc,LHCb:2025ray}. 
We highlight that the endpoint enhancement, tamed by the $k_T$ resummation effect, is crucial for explaining the scaling of $Q^4F_1(Q^2)$, which cannot be achieved in the QCDF \cite{Huang:2024ugd,Chen:2024fhj}.
Our work elaborates the necessity of the proton three-dimensional structure at high twists for understanding the $ep$ elastic scattering data. 

\textbf{\textit{Form factors and the pQCD framework}}--The matrix element of the electromagnetic current 
$j_\mu = \sum_{q=u,d} e_q {\bar q} \gamma_\mu q$ sandwiched between the proton states defines the Dirac FF $F_1(Q^2)$ and the Pauli FF $F_2(Q^2)$,  
\beq {\cal A}_\mu &\equiv& \langle P^\prime \vert j_\mu(0) \vert P \rangle \nonumber\\
&=& {\bar u}(P^\prime) \left[ \gamma_\mu F_1(Q^2) - i \frac{\sigma_{\mu\nu} q^\nu}{2m_p} F_2(Q^2) \right] u(P). 
\label{eq:EMFFs-definition} \eeq
Here $P$ and $P^\prime$ are the initial and final proton momenta satisfying the on-shell condition $P^2 = P^{\prime 2} = m_p^2$, 
$Q^2 \equiv -q^2$ with the momentum transfer $q = P - P^\prime$, and 
$\sigma_{\mu\nu} = i [\gamma_\mu, \gamma_\nu]/2$. The FFs at the full-recoiled point $Q^2 =0$ are normalized to the proton charge and magnetic moment,
$F_1(0) = 1$ and $F_1(0) + F_2(0) = 2.793$ \cite{PDG2024}.  

The momenta of the proton and of the valence quarks participating in the hard scattering are parametrized, in the Breit frame, as 
\beq &&P_\mu = \frac{1}{\sqrt{2}} \left( P^+ , P^-, {\bf 0} \right), \; 
k_{i \mu} = \left( x_i \frac{P^+}{\sqrt{2}}, 0, {\bf k}_{iT}\right), \nonumber\\
&&P^\prime_\mu = \frac{1}{\sqrt{2}} \left( P^-, P^+, {\bf 0} \right), \; 
k^\prime_{i \mu} = \left( 0, x^\prime_i \frac{P^{\prime -}}{\sqrt{2}}, {\bf k}^\prime_{iT}\right), 
\label{eq:kinematics}\eeq
in which the plus (minus) component reads $P^+ = E + Q/2$ ($P^- = E - Q/2$) with the proton energy $E = \sqrt{ Q^2/4+m_p^2}$, 
and the parton momentum fractions $x_i^{(\prime)}$ and transverse momenta ${\bf k}^{(\prime)}_{iT}$ obey the conservation $\sum_{i} x^{(\prime)}_i=1$ and $\sum_{i} {\bf k}^{(\prime)}_{iT} = 0$. 

The pQCD factorization formula for the $ep$ scattering amplitude is written as a convolution of the hard kernel ${\cal H}$ with the proton wave functions $\psi$,
\beq && {\cal A}_\mu = \sum_{n=1}^{42} \sum_{t,t^\prime = 3}^6 \int [dx] [dx^{\prime}] [d^2 {\bf b}] [d^2 {\bf b}^{\prime}]
\, \psi^{t^\prime}(x^\prime_i, b^\prime_i, Q, \mu_f) \nonumber\\ 
&& \hspace{9mm} {\cal H}^{t,t^\prime}_{n, \mu}(x_i,x^\prime_i, b_{i}, b^\prime_{i}, Q, \mu_r, \mu_f)  \psi^{t}(x_i, b_i, Q, \mu_f), 
\label{eq:ffs-pQCD} \eeq 
$[dx^{(\prime)}] = dx_1^{(\prime)} dx_2^{(\prime)} dx_3^{(\prime)} \delta(1-x_1^{(\prime)}-x_2^{(\prime)}-x_3^{(\prime)})$ being the longitudinal phase-space measure.
The measure  $[d^2 {\bf b}^{(\prime)}]$ is defined according to the Fourier transformation from the $k^{(\prime)}_T$ space to the impact parameter space 
with the integration intervals $ b^{(\prime)}_{i}\in [1/Q, 1/\Lambda_{\rm QCD}]$, $\Lambda_{\rm QCD}$ being the QCD scale. The index $n$ enumerates all $42$ leading-order (LO) Feynman diagrams, 
$t$ ($t^{\prime}$) labels the twists 3-6 of the initial (final) proton LCDAs, and $\mu_r$ ($\mu_f$) represents the renormalization (factorization) scale. 

The parton transverse momenta span three hierarchical scales: $Q$  (hard), $\sqrt{Q\Lambda_{\rm QCD}}$ (hard-collinear) and $\Lambda_{\rm QCD}$ (soft). 
The infrared logarithms associated with the soft scale are absorbed into the wave functions $\psi$. 
The large logarithms associated with the hard-collinear scale are resummed into the well-known Sudakov factors in the impact parameter space, and factored out of $\psi$, $\psi = e^{-S(x_i, b_i, Q)} \phi(x_i, b_i,\mu_f) $, $\phi$ being the proton LCDAs.
The explicit expression for the exponent $S$ have been given in Refs. \cite{Li:1992ce,Botts:1989kf,Li:1992nu}.
The scales $\mu_r=\mu_f$ for each diagram are set to the maximal virtuality of internal quarks and gluons. 
With the soft and hard-collinear dynamics being organized appropriately, Eq. (\ref{eq:ffs-pQCD}) is applicable to exclusive processes at intermediate energies; 
the low $k_T$ (high $b$) region would be strongly suppressed by the Sudakov factors, 
such that internal particle virtualities are governed by finite transverse momenta in the endpoint region. 
We will ignore the intrinsic $b$ dependence \cite{Chai:2024tss} and the factorization scale $\mu_f$ in $\phi(x_i, b_i,\mu_f) $ for simplicity, 
so that the parton transverse distributions are completely described by the Sudakov factors.

\textbf{\textit{Twist expansion and LCDAs}}--The proton-to-vacuum matrix elements define eight independent LCDAs: 
one at twist $3$, three at twists $4$ and $5$, and one at twist $6$. 
These LCDAs admit a conformal expansion, with eight nonperturbative parameters specified by isospin symmetry and equations of motion. 
For instance, the dimensionful parameters $f_p$, $\lambda_1$ and $\lambda_2$ are fixed 
by the local matrix elements at LO in the conformal spin expansion \cite{Braun:2000kw}, 
\beq &&\langle 0 \vert \varepsilon^{ijk} \left[ u^i C \zsl u^j \right] \gamma_5 \zsl d^k\vert P \rangle = f_p P^+  \zsl u(P), \nonumber\\
&&\langle 0 \vert \varepsilon^{ijk} \left[ u^i C \gamma_\mu u^j \right] \gamma_5 \gamma^\mu d^k\vert P \rangle = \lambda_1 m_p u(P), \nonumber\\
&&\langle 0 \vert \varepsilon^{ijk} \left[ u^i C \sigma_{\mu\nu} u^j \right] \gamma_5 \sigma^{\mu\nu} d^k \vert P \rangle = \lambda_2 m_p u(P), \label{eq:paras-LCP}\eeq
respectively, where $C$ stands for the charge conjugation, and $z$ is a light-like vector chosen in the minus direction. 
At NLO of the conformal spin expansion, five additional dimensionless parameters are related to local operators containing one derivative: 
$V_1^d$, $f_1^d$, $f_2^d$ ($A_1^u$, $f_1^u$) are associated with the derivatives on the $d$-quark ($u$-quark) fields in Eq. (\ref{eq:paras-LCP}). 
The precise information on the above parameters are not yet available. 
Results from QCDSR \cite{Braun:2000kw,Braun:2006hz} and LQCD \cite{RQCD:2019hps}, both acquired by the Regensburg group, 
exhibit magnificent discrepancies and uncertainties, even for the LO conformal spin parameters. Alternative methods for determining these parameters, such as data-driven analyses of the FFs, 
are worthwhile and can yield critical insights into the proton structure. 

\begin{table}[t]\vspace{-4mm}
\caption{Typical integrands of $F_1(Q^2)$ from each combination of the initial (column) and final (row) proton LCDAs of various twists.} \vspace{2mm}
\renewcommand{\arraystretch}{2.0} 
\setlength{\tabcolsep}{2pt}       
\begin{tabular}{ c | c | c | c | c }
\toprule[1pt]
\textbf{Twists} & $3$ & $4$ & $5$ & $6$ \\ \hline
$3$ & $\frac{[x_i]^3[x^\prime_i]^3f_p^2}{\tilde{Q}^{4}}$ & $\frac{[x_i]^3[x^\prime_i]^2f_p \lambda_j}{\tilde{Q}^{5}}$  & 
$\frac{[x_i]^3 [x^\prime_i] f_p \lambda_j}{\tilde{Q}^{6}}$  & $\frac{[x_i]^3 f_p^2}{\tilde{Q}^{7}}$   \\ \hline
$4$ & $\frac{[x_i]^2 [x^\prime_i]^3 \lambda_j f_p}{\tilde{Q}^{5}}$  & $\frac{[x_i]^2 [x^\prime_i]^2 [\lambda_j]^2}{\tilde{Q}^{6}}$  & 
$\frac{[x_i]^2 [x^\prime_i] [\lambda_j]^2}{\tilde{Q}^{7}}$  & $\frac{[x_i]^2 \lambda_j f_p}{\tilde{Q}^{8}}$   \\ \hline
$5$ & $\frac{[x_i] [x^\prime_i]^3 \lambda_j f_p }{\tilde{Q}^{6}}$ & $\frac{[x_i] [x^\prime_i]^2 [\lambda_j]^2 }{\tilde{Q}^{7}}$  & 
$\frac{[x_i] [x^\prime_i] [\lambda_j]^2}{\tilde{Q}^{8}}$  & $\frac{[x_i] \lambda_j f_p}{\tilde{Q}^{9}}$   \\ \hline
 $6$ & $\frac{[x^\prime_i]^3 f_p^2}{\tilde{Q}^{7}}$  & $\frac{[x^\prime_i]^2 f_p \lambda_j}{\tilde{Q}^{8}}$  & $\frac{[x^\prime_i] f_p \lambda_j}{\tilde{Q}^{9}}$  & $\frac{f_p^2}{\tilde{Q}^{10}}$   \\ \hline
\toprule[1pt]
\end{tabular} \vspace{-2mm} \label{tab:ff-power}
\end{table}

The typical integrands for $F_1(Q^2)$ from each combination of the initial (column) and final (row) proton LCDAs of various twists 
are collected in Table \ref{tab:ff-power}, where the naive counting of the powers in the longitudinal virtuality 
$\tilde{Q}^2\sim x_ix'_jQ^2$, $x_iQ^2$ or $x'_iQ^2$ is displayed. 
The FF scales like ${\cal O}(1/\tilde{Q}^4)$ asymptotically, which is corrected by higher-twist contributions of ${\cal O}(1/\tilde{Q}^{n-2})$, $n = t+t^\prime$. 
The twist-$3$ LCDAs are tripartite functions with three-quark correlations, the twist-$4$ and twist-$5$ LCDAs reduce to bipartite and linear functions, respectively, and the twist-6 LCDAs are independent of the momentum fractions. 
The above explicates the powers of the momentum fractions in the numerators, and the increasing importance of the endpoint regions with twists. 
The twist-$3$ and -$6$ LCDAs are proportional to $f_p$, and the twist-$4$ and -$5$ LCDAs are proportional to $\lambda_i$, 
which are an order of magnitude larger than $f_p$. 
Hence, higher-twist contributions may not be suppressed numerically relative to the leading-twist one at intermediate $Q^2$.

\begin{figure}[b] 
\begin{center} 
\includegraphics[width=0.8\linewidth]{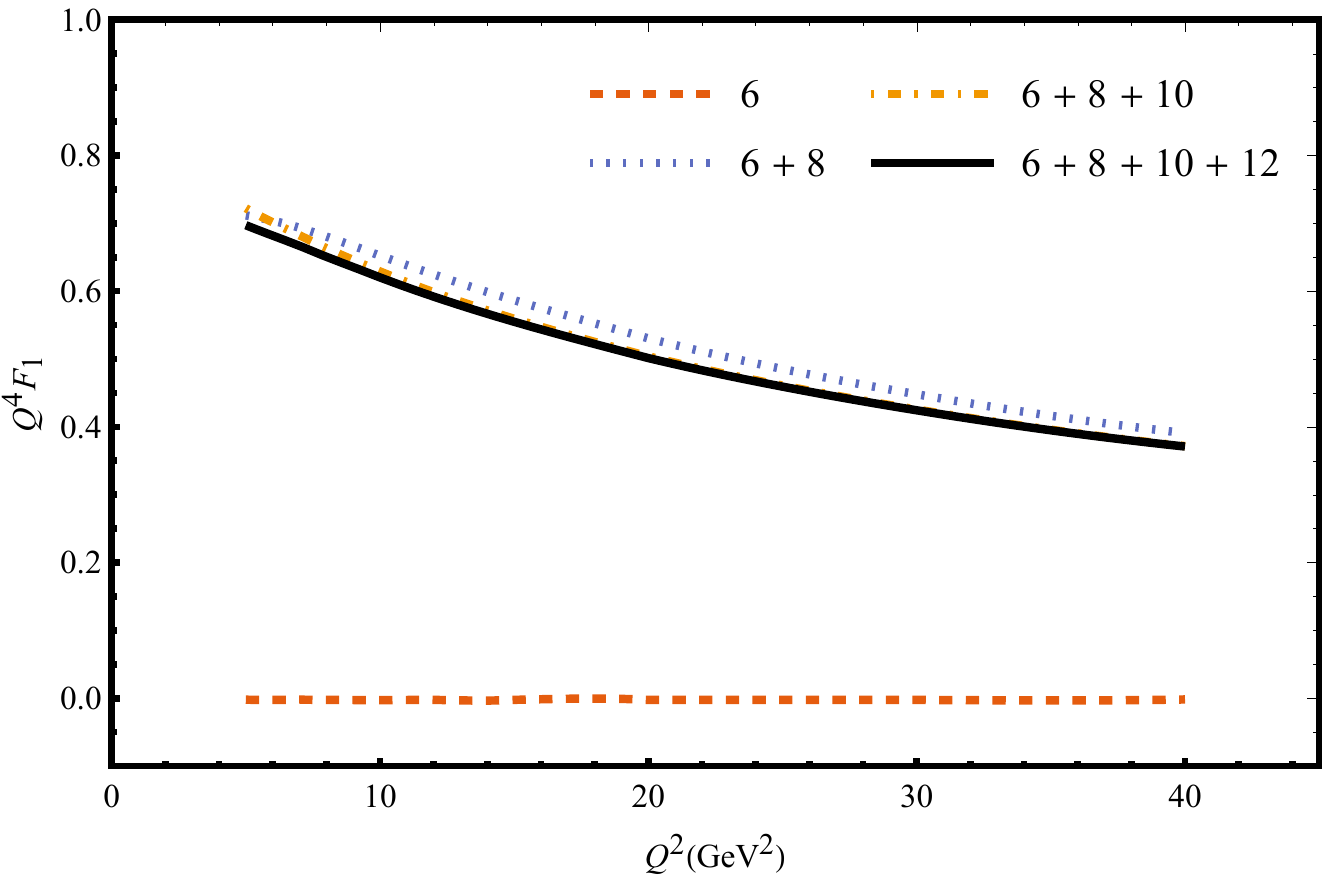} 
\end{center} \vspace{-4mm} 
\caption{Contributions of different twists $n = t + t^\prime$ to the Dirac FF $Q^4F_1(Q^2)$ with the QCDSR parameters \cite{Braun:2006hz}.} \vspace{-2mm} 
\label{fig:twist} 
\end{figure}

\begin{figure*}[t]
\begin{center} \vspace{4mm}
\includegraphics[width=0.3\textwidth]{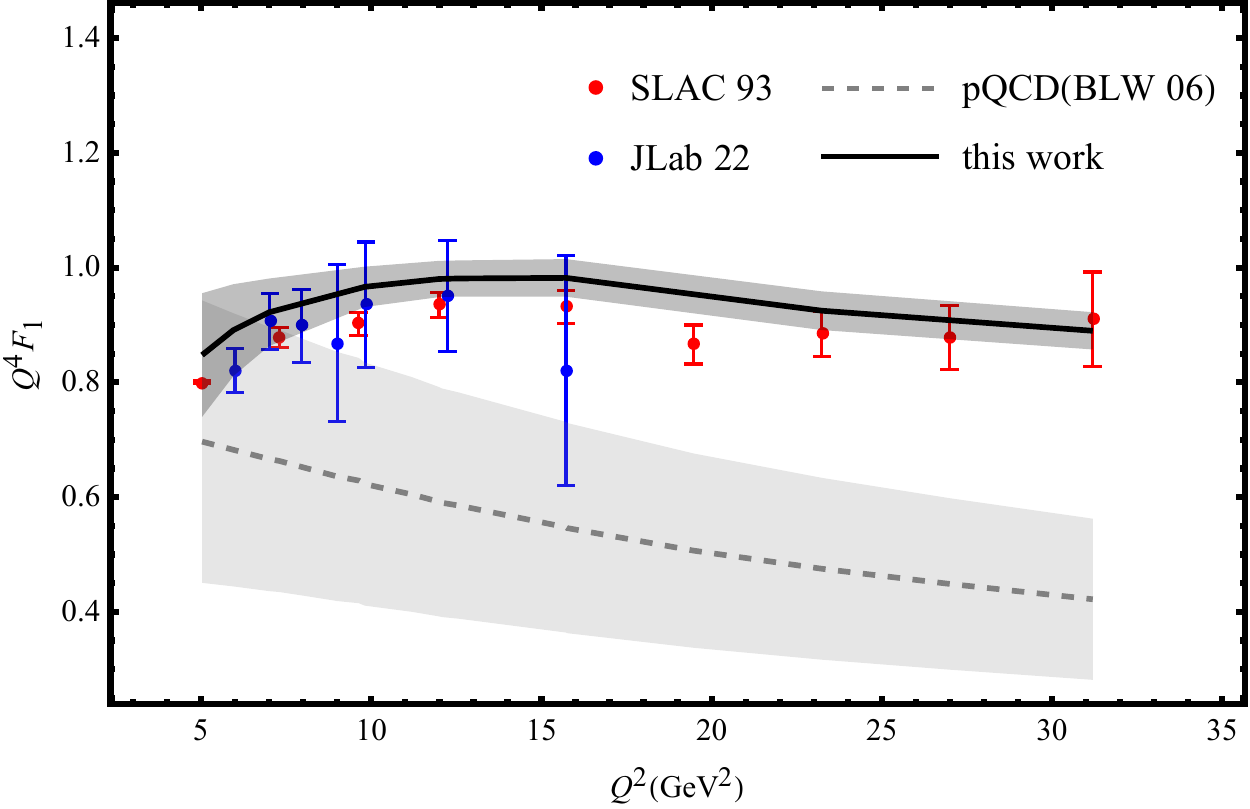} 
\hspace{2mm}
\includegraphics[width=0.3\textwidth]{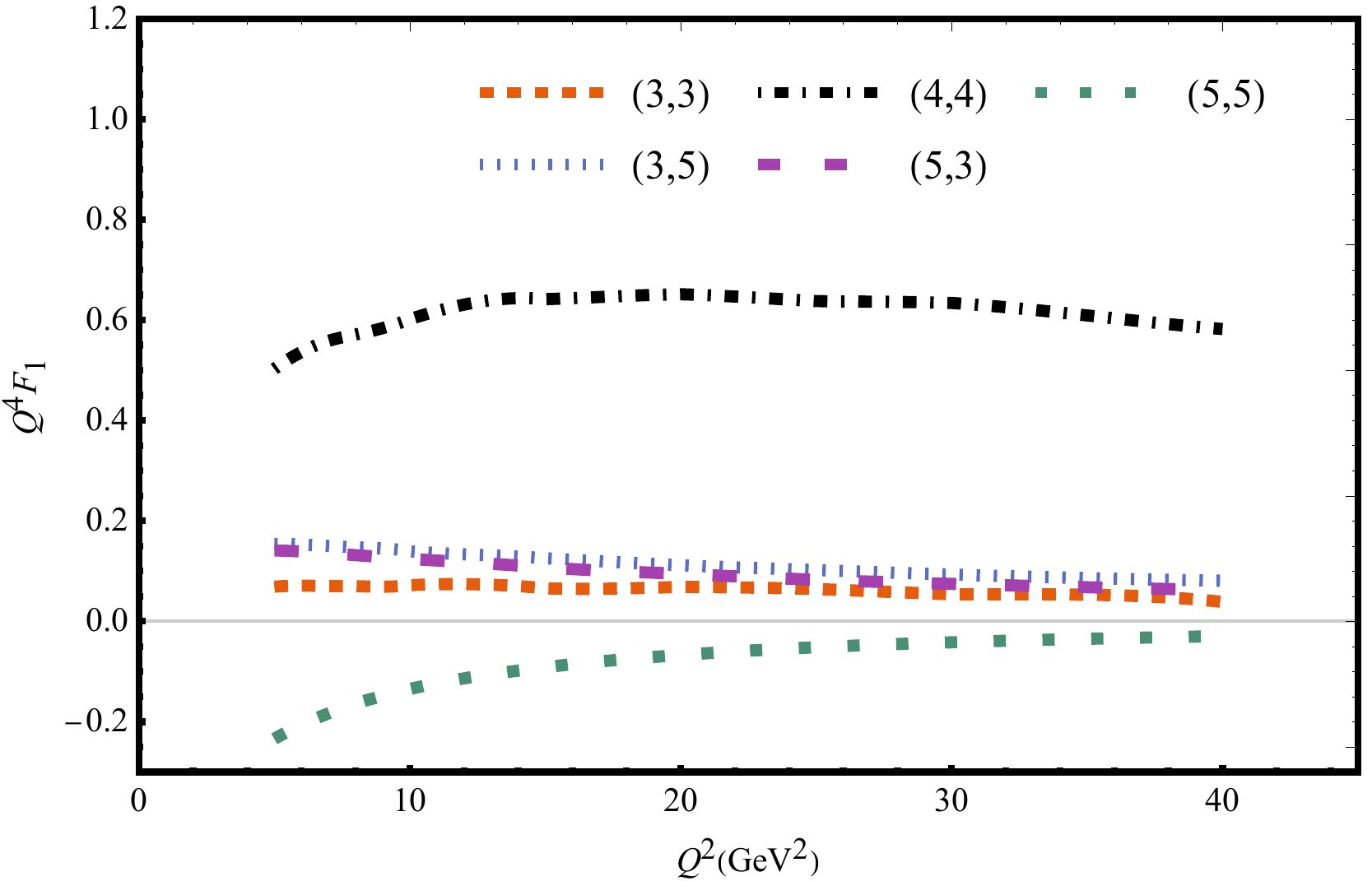}\hspace{2mm}
\includegraphics[width=0.3\textwidth]{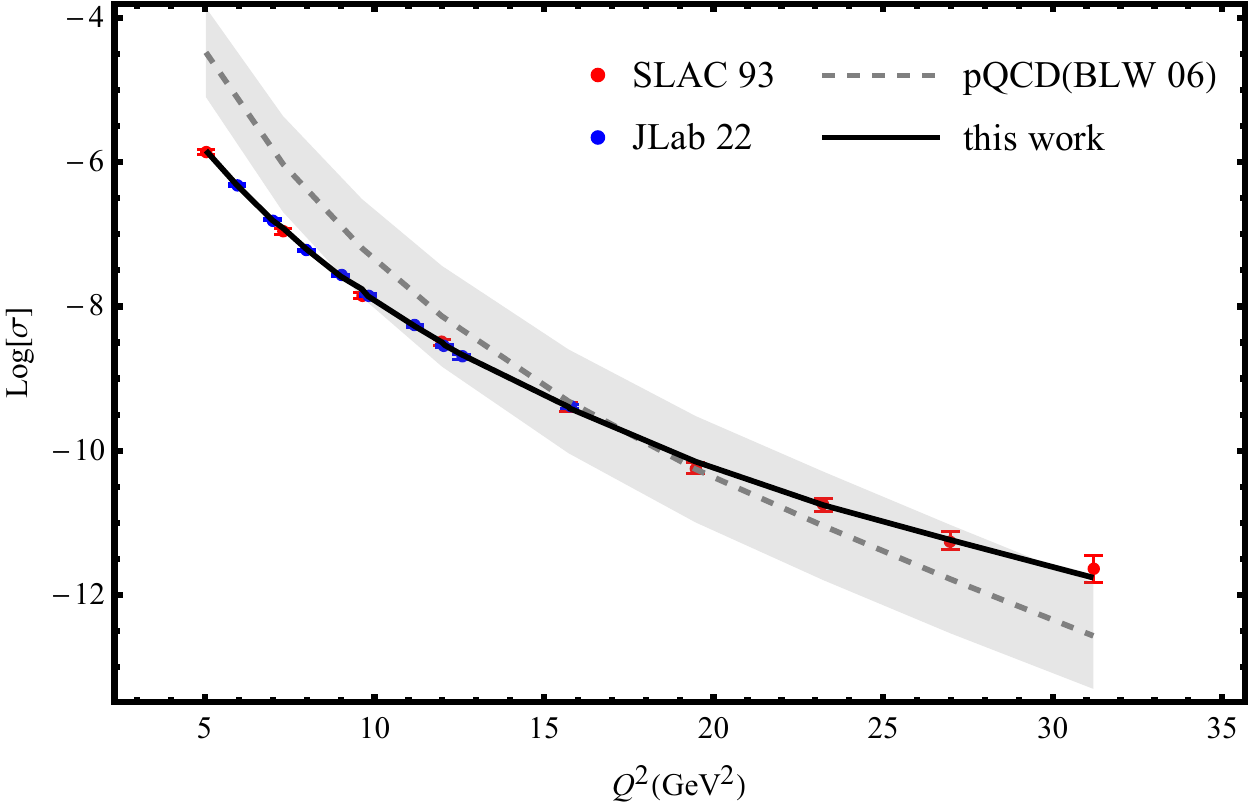}\vspace{-4mm}
\end{center} 
$(a)$ \hspace{5.5cm}(b)\hspace{5.5cm}(c)
\caption{pQCD predictions for (a) the Dirac FF $Q^4F_1(Q^2)$, (b) $Q^4F_1(Q^2)$ from each twist combination $(t,t')$, and (c) $ep$ elastic scattering cross section, where "BLW 06" refers to the input of the QCDSR parameters \cite{Braun:2006hz}.}
\label{fig:result-data}
\end{figure*}

Figure \ref{fig:twist} presents the $Q^2$ dependencies of the various twist contributions to $Q^4F_1(Q^2)$, as calculated using the QCDSR inputs for the nonperturbative parameters \cite{Braun:2006hz}. For simplicity, only contributions from even twists are displayed. It turns out that  the pieces from twist $n=8$ dominate, and the power expansion converges above $n=9$.  
Comparing Fig. \ref{fig:twist} with the corresponding data in Fig. \ref{fig:result-data}(a), 
we see that the predicted $Q^2$ dependence deviates dramatically from the observed scaling within $5 \leq Q^2 \leq 30$ GeV$^2$, 
though the magnitudes of $Q^4F_1(Q^2)$ are of the same order. 
We stress that the roughly constant $Q^4F_1(Q^2)$ around intermediate $Q^2$ cannot be understood in collinear factorization, 
since the dominant $1/\tilde{Q}^8$ contribution leads to the descent of $Q^4F_1(Q^2)$ with $Q^2$ definitely. Considering  the additional hard-collinear scale $k_T$, which may modify the power-law behaviors associated with different twists, we examine this issue in the pQCD approach.

The aforementioned puzzle invokes a data-driven analysis on the proton FFs in view of the uncertain LCDA inputs. 
Introducing a set of eight free parameters $\left[ a_i \right] = \{f_p, f_p A_1^u,f_p V_1^d, \lambda_1, \lambda_1 f_1^d, 
\lambda_1 f_1^u, \lambda_2, \lambda_2f_2^d \}$, we express the pQCD formulas for the proton Dirac and Pauli FFs as 
\beq F(Q^2) = \left[ a_i \right]^T 
{\cal H}_{ij}(Q^2) \left[ a_j \right]. 
\label{eq:ff-para}\eeq
If two parameters are always present as a product in the formula, the product is chosen as an element of $\left[ a_i \right]$.
The hard matrix ${\cal H}_{ij}$ depends only on $Q^2$, which has absorbed the scale evolution of the parameters. The pQCD predictions for the reduced $ep$ cross section at selected $Q_l^2$ points $\sigma_l^{\rm pQCD} \equiv \sigma^{\rm pQCD}(Q_l^2)$ are obtained by inserting Eq. (\ref{eq:ff-para}) into \cite{Christy:2021snt}
\beq \sigma= \frac{Q^2}{4m_p^2} \left( F_1+F_2 \right)^2+ \epsilon \left( F_1- \frac{Q^2}{4m_p^2} F_2 \right)^2, \label{eq:cross-section} \eeq 
with $\epsilon = \left[ 1 + 2 \left(1 + Q^2/(4m_p^2) \right) \tan^2(\theta/2) \right]^{-1}$, $\theta$ being the scattering angle. 

It is challenging to constrain the overall parameters $f_p$, $\lambda_1$ and $\lambda_2$ in the LCDAs from experimental data alone. 
Since they are known from LQCD with uncertainties all below $10\%$ \cite{RQCD:2019hps}, we incorporate them as priors in our fit to the measured $ep$ cross section. 
The eight parameters are then fixed by minimizing 
\beq \chi^2 = \sum_{l=1}^{18} \frac{(\sigma^{\rm pQCD}_l-\sigma^{\rm data}_l)^2}{(\Delta\sigma^{\rm data}_l)^2} + \sum_{k=1}^{3} \frac{(a_k-a_k^{\rm LQCD})^2}{(\Delta a_k^{\rm LQCD})^2}
\label{eq:fit-chi}, \eeq
where $\Delta \sigma^{\rm data}$ ($\Delta a_k^{\rm LQCD}$) denotes the uncertainty of the data $\sigma^{\rm data}_l$ at $Q_l^2$ (the lattice value $a_k^{\rm LQCD}$), and the index $k$ runs through the prior settings for $f_p$, $\lambda_1$, and $\lambda_2$. We take the eighteen data sets from SLAC \cite{Sill:1992qw} and JLab \cite{Christy:2021snt} at high $Q^2$ and corresponding scattering angles, for which the pQCD evaluations are reliable.   

\begin{table*}[t] \centering
\caption{Parameters in the proton LCDAs from our fit, QCDSR \cite{Braun:2006hz} and LQCD \cite{RQCD:2019hps}, where $f_p$, $\lambda_1$ and $\lambda_2$ are given in unit of GeV$^2$.} \vspace{2mm}
\renewcommand{\arraystretch}{1.2}  
\begin{tabular*}{180mm}{l@{\extracolsep{\fill}}cccc cccc}
\toprule[1pt]
&$f_N$ ($10^{-3}$) & $\lambda_1$ ($10^{-2}$) & 
$\lambda_2$ ($10^{-2}$) & $V_1^d$ & 
$A_1^u$ & $f_1^d$ & $f_2^d$ & $f_1^u$ \\ \hline
QCDSR & $5.0\pm0.5$ & 
$-2.7\pm0.9$ & $5.4\pm1.9$ & $0.23\pm0.03$ 
& $0.38\pm0.15$ & $0.4\pm0.05$ & $0.22\pm0.05$ & $0.07\pm0.05$ \\
LQCD & $3.67\pm0.06$ & $-4.02\pm0.38$ & $8.37\pm0.43$ & $0.288\pm0.007$ 
& $0.096\pm0.010$ &...&...&... \\
this work & $3.67\pm0.06$ & $-3.98\pm0.37$ 
& $8.32\pm0.43$ & $1.15\pm0.44$
& $1.27\pm0.72$ & $0.85\pm0.12$ 
& $0.59\pm0.04$ & $-0.23\pm0.03$ \\ 
\toprule[1pt] \end{tabular*}
\label{tab:LCDAspara}
\end{table*}

The parameters from the fit with $\chi^2/d.o.f.=1.6$ are compared with those from QCDSR \cite{Braun:2006hz} and LQCD \cite{RQCD:2019hps} in Table \ref{tab:LCDAspara}.
The fit stabilizes the prior set of $f_p$, $\lambda_1$ and $\lambda_2$ derived in LQCD, 
including their uncertainties, confirming that these values are reasonable. 
However, significant shifts in $V_1^d$, $A_1^u$, $f_1^u$, $f_1^d$ and $f_2^d$ relative to QCDSR and LQCD results arppear. Because our pQCD study is conducted at LO in $\alpha_s$, the distinction can be conclusive after higher-order corrections are taken into account. Some extracted parameters come with large uncertainties, which are due to the strong parameter correlations, 
though the underlying data are very precise. 
The predictions from the fitted LCDAs for $Q^4F_1(Q^2)$ are confronted with those from the QCDSR parameters \cite{Braun:2006hz} in Fig.~\ref{fig:result-data}(a). 
The former feature well-controlled uncertainties and a mild $Q^2$ dependence across the intermediate-to-large $Q^2$, 
in sharp contrast to the declining trends of the latter and in the previous QCDF calculations \cite{Huang:2024ugd,Chen:2024fhj}.
It is obvious that the agreement between the theory and the data \cite{Sill:1992qw,Christy:2021snt} has been greatly improved.

A closer look tells that the fitted LCDAs amplify more the endpoint contributions than those from QCDSR do.
When the endpoint region is further emphasized, one of the virtualities $\tilde Q^2\sim x_i x_j Q^2$ becomes small enough and negligible relative to the transverse momentum squared $k_T^2$, which is of order of the hard-collinear scale $Q\Lambda_{\rm QCD}$ under the Sudakov suppression. 
The other virtualities like $x_iQ^2$ remain larger than $k_T^2$ in this configuration.
Once the replacement by $k_T^2$ occurs, the dominant twist-8 contribution from the combination $(t,t')=(4,4)$ scales like $1/Q^4$ effectively, and $Q^4F_1(Q^2)$ turns into a constant
as depicted in Fig. \ref{fig:result-data}(b). When $Q^2$ keeps growing, $\tilde Q^2$ dictates eventually, and $Q^4F_1(Q^2)$ starts to descend. 
Our numerical inspection shows that $Q^4F_1(Q^2)$ decreases by half at $Q^2\sim 150$ GeV$^2$.
We advocate that the scaling of $Q^4F_1(Q^2)$ can be reproduced only in $k_T$ factorization, but not in collinear factorization.
In this sense, the flat $Q^4F_1(Q^2)$ around intermediate $Q^2$ has evinced the relevance of parton transverse degrees of freedom in hard exclusive processes, for which three-dimension structures of baryons serve as essential inputs.

Figure \ref{fig:result-data}(c) indicates that the pQCD prediction for the $ep$ cross section $\sigma$ matches perfectly the data in the range $5 \leq Q^2 \leq 30$ GeV$^2$. The outcome from the QCDSR parameters fails to account for the observed $Q^2$ dependence apparently. 
At last, we verify the determined LCDAs by applying them to update the pQCD predictions for the $\Lambda_b \to pK$ decay. Based on their universality, we obtain a branching ratio of $Br=(4.4^{+2.6}_{-1.1})\times 10^{-6}$ and a direct CP asymmetry of $A_{CP}=(3^{+2}_{-1}) \%$. The quoted uncertainties reflect only the dominant error source, which originates from the $\Lambda_b$ baryon LCDAs. 
Compared with the previous pQCD results, which relied on QCDSR parameters and yielded $Br=(2.8^{+3.3}_{-1.6})\times 10^{-6}$ and $A_{CP}=(-6^{+3}_{-2})\%$ \cite{Han:2024kgz}, the updated calculations reveal better consistency with the experimental data $Br=(5.5\pm1.0)\times 10^{-6}$ and $A_{CP}=(-1.1\pm0.8)\%$ \cite{LHCb:2024iis}. 

\textbf{\textit{Conclusion}}--We have investigated the proton FFs in the pQCD $k_T$ factorization framework, and uncovered several important competing mechanisms related to the proton three-dimensional  structures: the endpoint enhancement in hard kernels, the power suppression by internal particle virtualities associated with higher-twist LCDAs, and the modulation of hard-collinear transverse momenta by the Sudakov resummation. 
The study demonstrates the delicate interplay among the different mechanisms for accommodating the unique scaling of $Q^4F_1(Q^2)$ in the region of intermediate $Q^2$, which cannot be realized in collinear factorization.
Taking LQCD values as priors, we extracted the nonperturbative LCDA parameters from the fits to the measured $ep$ elastic scattering cross section. 
It is expected that the inclusion of higher-order corrections may help resolve the discrepancies between our fits and those from LQCD and QCDSR, and improve more the agreement between the predictions and the data for heavy baryon decays.
Our work highlights the crucial role of baryon three-dimensional  structures in factorization-based approaches to exclusive processes, and sets the precision requirement for their reliable perturbative analyses.

{\it Acknowledgments:} We are grateful to Long-bin Chen, Ji-feng Hu, Guang-shun Huang and Yu-ming Wang for fruitful discussions. 
This work is supported by the National Key R$\&$D Program of China under Contract No. 2023YFA1606000, the National Science Foundation of China (NSFC) under Grant No. 12575098 and No. 12335003 and the Fundamental Research Funds for the Central Universities (lzujbky-2023-stlt01).

\newpage


\begin{thebibliography}{99}

\bibitem{EuropeanMuon:1987isl}
J.~Ashman \textit{et al.} [European Muon],
Phys. Lett. B \textbf{206} (1988), 364.

\bibitem{Kuhn:2008sy}
S.~E.~Kuhn, J.~P.~Chen and E.~Leader,
Prog. Part. Nucl. Phys. \textbf{63} (2009), 1-50.

\bibitem{Ji:2020ena}
X.~Ji, F.~Yuan and Y.~Zhao,
Nature Rev. Phys. \textbf{3} (2021) no.1, 27-38.

\bibitem{Yang:2016plb}
Y.~B.~Yang, R.~S.~Sufian, A.~Alexandru, T.~Draper, M.~J.~Glatzmaier, K.~F.~Liu and Y.~Zhao,
Phys. Rev. Lett. \textbf{118} (2017) no.10, 102001.

\bibitem{Ji:1994av}
X.~D.~Ji,
Phys. Rev. Lett. \textbf{74} (1995), 1071-1074.

\bibitem{Lorce:2017xzd}
C.~Lorc\'e,
Eur. Phys. J. C \textbf{78} (2018) no.2, 120.

\bibitem{Yang:2018nqn}
Y.~B.~Yang, J.~Liang, Y.~J.~Bi, Y.~Chen, T.~Draper, K.~F.~Liu and Z.~Liu,
Phys. Rev. Lett. \textbf{121} (2018) no.21, 212001.

\bibitem{Pohl:2010zza}
R.~Pohl, A.~Antognini, F.~Nez, F.~D.~Amaro, F.~Biraben, J.~M.~R.~Cardoso, D.~S.~Covita, A.~Dax, S.~Dhawan and L.~M.~P.~Fernandes, \textit{et al.}
Nature \textbf{466} (2010), 213-216.

\bibitem{Antognini:2013txn}
A.~Antognini, F.~Nez, K.~Schuhmann, F.~D.~Amaro, FrancoisBiraben, J.~M.~R.~Cardoso, D.~S.~Covita, A.~Dax, S.~Dhawan and M.~Diepold, \textit{et al.}
Science \textbf{339} (2013), 417-420.

\bibitem{Gao:2021sml}
H.~Gao and M.~Vanderhaeghen,
Rev. Mod. Phys. \textbf{94} (2022) no.1, 015002.

\bibitem{Fischer:2004jt}
M.~Fischer, N.~Kolachevsky, M.~Zimmermann, R.~Holzwarth, T.~Udem, T.~W.~Hansch, M.~Abgrall, J.~Grunert, I.~Maksimovic and S.~Bize, \textit{et al.}
Phys. Rev. Lett. \textbf{92} (2004), 230802.

\bibitem{Karshenboim:2005iy}
S.~G.~Karshenboim,
Phys. Rept. \textbf{422} (2005), 1-63.

\bibitem{Hofstadter:1953zjy}
R.~Hofstadter, H.~R.~Fechter and J.~A.~McIntyre,
Phys. Rev. \textbf{92} (1953) no.4, 978.

\bibitem{Hofstadter:1956qs}
R.~Hofstadter,
Rev. Mod. Phys. \textbf{28} (1956), 214-254.

\bibitem{Lepage:1979za}
G.~P.~Lepage and S.~J.~Brodsky,
Phys. Rev. Lett. \textbf{43} (1979) no.21, 545-549 [erratum: Phys. Rev. Lett. \textbf{43} (1979), 1625-1626]; 
Phys. Lett. B \textbf{87} (1979), 359-365; 
Phys. Rev. D \textbf{22} (1980), 2157.

\bibitem{Chernyak:1977fk}
V.~L.~Chernyak, A.~R.~Zhitnitsky and V.~G.~Serbo, JETP Lett. \textbf{26} (1977), 594-597;
A.~V.~Efremov and A.~V.~Radyushkin, Phys. Lett. B \textbf{94} (1980), 245-250;
V.~L.~Chernyak and A.~R.~Zhitnitsky, Phys. Rept. \textbf{112} (1984), 173.

\bibitem{Chernyak:1984bm}
V.~L.~Chernyak and I.~R.~Zhitnitsky,
Nucl. Phys. B \textbf{246} (1984), 52-74.

\bibitem{Ji:1986uh}
C.~R.~Ji, A.~F.~Sill and R.~M.~Lombard,
Phys. Rev. D \textbf{36} (1987), 165.

\bibitem{Carlson:1987sw}
C.~E.~Carlson and F.~Gross,
Phys. Rev. D \textbf{36} (1987), 2060.

\bibitem{Thomson:2006ny}
R.~Thomson, A.~Pang and C.~R.~Ji,
Phys. Rev. D \textbf{73} (2006), 054023.

\bibitem{Brodsky:1973kr}
S.~J.~Brodsky and G.~R.~Farrar,
Phys. Rev. Lett. \textbf{31} (1973), 1153-1156.

\bibitem{Mueller:1981sg}
A.~H.~Mueller,
Phys. Rept. \textbf{73} (1981), 237.

\bibitem{SL-pQCD}
S. J. Brodsky and G. P. Lepage, in Perturbative Quantum Chromodynamics, edited by A.H. Mueller (World Scientific, Singapore, 1989).


\bibitem{Chen:2024fhj}
L.~B.~Chen, W.~Chen, F.~Feng, S.~Hu and Y.~Jia,
Phys. Rev. Lett. \textbf{135} (2025) no.13, 131903.

\bibitem{Huang:2024ugd}
Y.~K.~Huang, B.~X.~Shi, Y.~M.~Wang and X.~C.~Zhao,
Phys. Rev. Lett. \textbf{135} (2025) no.6, 061901.

\bibitem{Li:1992ce}
H.~n.~Li,
Phys. Rev. D \textbf{48} (1993), 4243-4254.

\bibitem{Chai:2024tss}
J.~Chai and S.~Cheng,
Phys. Rev. D \textbf{111} (2025) no.7, L071902.

\bibitem{Arnold:1986nq}
R.~G.~Arnold, P.~E.~Bosted, C.~C.~Chang, J.~Gomez, A.~T.~Katramatou, C.~J.~Martoff, G.~Petratos, A.~A.~Rahbar, S.~Rock and A.~F.~Sill, \textit{et al.}
Phys. Rev. Lett. \textbf{57} (1986), 174.

\bibitem{Sill:1992qw}
A.~F.~Sill, R.~G.~Arnold, P.~E.~Bosted, C.~C.~Chang, J.~Gomez, A.~T.~Katramatou, C.~J.~Martoff, G.~Petratos, A.~A.~Rahbar and S.~Rock, \textit{et al.}
Phys. Rev. D \textbf{48} (1993), 29-55.

\bibitem{Rock:1991jy}
S.~Rock, R.~G.~Arnold, P.~E.~Bosted, B.~T.~Chertok, B.~A.~Mecking, I.~A.~Schmidt, Z.~M.~Szalata, R.~York and R.~Zdarko,
Phys. Rev. D \textbf{46} (1992), 24-44.

\bibitem{Christy:2021snt}
M.~E.~Christy, T.~Gautam, L.~Ou, B.~Schmookler, Y.~Wang, D.~Adikaram, Z.~Ahmed, H.~Albataineh, S.~F.~Ali and B.~Aljawrneh, \textit{et al.}
Phys. Rev. Lett. \textbf{128} (2022) no.10, 102002.

\bibitem{Chernyak:1987nt}
V.~L.~Chernyak, A.~A.~Ogloblin and I.~R.~Zhitnitsky,
Sov. J. Nucl. Phys. \textbf{48} (1988), 536.

\bibitem{RQCD:2019hps}
G.~S.~Bali \textit{et al.} [RQCD],
Eur. Phys. J. A \textbf{55} (2019) no.7, 116.

\bibitem{Anikin:2013aka}
I.~V.~Anikin, V.~M.~Braun and N.~Offen,
Phys. Rev. D \textbf{88} (2013), 114021.

\bibitem{Han:2024kgz}
J.~J.~Han, J.~X.~Yu, Y.~Li, H.~n.~Li, J.~P.~Wang, Z.~J.~Xiao and F.~S.~Yu,
Phys. Rev. Lett. \textbf{134} (2025) no.22, 221801.

\bibitem{Han:2025tvc}
J.~J.~Han, J.~X.~Yu, Y.~Li, H.~n.~Li, J.~P.~Wang, Z.~J.~Xiao and F.~S.~Yu,
Phys. Rev. D \textbf{112} (2025) no.5, 053007.

\bibitem{LHCb:2025ray}
R.~Aaij \textit{et al.} [LHCb],
Nature \textbf{643} (2025) no.8074, 1223-1228.

\bibitem{PDG2024}
S. Navas et al. (Particle Data Group), Phys. Rev. D 110, 030001 (2024) and 2025 update.

\bibitem{Botts:1989kf}
J.~Botts and G.~F.~Sterman,
Nucl. Phys. B \textbf{325} (1989), 62-100.

\bibitem{Li:1992nu}
H.~n.~Li and G.~F.~Sterman,
Nucl. Phys. B \textbf{381} (1992), 129-140.

\bibitem{Braun:2006hz}
V.~M.~Braun, A.~Lenz and M.~Wittmann,
Phys. Rev. D \textbf{73} (2006), 094019.

\bibitem{Braun:2000kw}
V.~Braun, R.~J.~Fries, N.~Mahnke and E.~Stein,
Nucl. Phys. B \textbf{589} (2000), 381-409 [erratum: Nucl. Phys. B \textbf{607} (2001), 433-433].


\bibitem{LHCb:2024iis}
R.~Aaij \textit{et al.} (LHCb Collaboration),
Phys. Rev. D \textbf{111}, 092004 (2025).

\end{thebibliography}
\end{document}